\documentclass[runningheads]{llncs}

\usepackage{booktabs}
\usepackage{graphicx}
\usepackage[table]{xcolor}
\usepackage[labelfont=normalfont]{subcaption}
\usepackage[labelfont+=bf,labelsep=period]{caption}
\usepackage{textcomp}
\usepackage{listings}
\usepackage[]{longtable}
\usepackage{tikz}
\usepackage{array}
\usepackage{makecell}
\usepackage{enumitem}
\usepackage{pgfplots}
\usepackage{multirow}
\usepackage[outline]{contour}
\usepackage{boldline}
\usepackage{tablefootnote}
\usepackage{footmisc}
\usepackage{threeparttable}

\usetikzlibrary{shapes.arrows}

\pgfplotsset{compat=1.15}
\usetikzlibrary{automata,arrows,positioning,petri}
\newcolumntype{L}[1]{>{\raggedright\arraybackslash }m{#1}}

\def\checkmark{\tikz\fill[scale=0.4](0,.35) -- (.25,0) -- (1,.7) -- (.25,.15) -- cycle;}
\def\halfcheck{\tikz\draw[scale=0.4](0,0) -- (0.5,0) -- (0.5,0.5) -- (0,.5) -- cycle;}
\def\cross{\tikz\draw[scale=0.15,line cap=round](0,0) -- (1,1) (0,1) -- (1,0);}
\newcolumntype{v}[1]{>{\centering\arraybackslash\setlength\baselineskip{0.8\baselineskip}}m{#1}}
\newcommand\NA{--}
\newcommand\setrowfont[1]{\noalign{\gdef\rfont{#1}}}

\PassOptionsToPackage{hyphens}{url}
\usepackage[hidelinks]{hyperref}



\begin{document}

\title{Challenges in Digital Twin Development for Cyber-Physical Production Systems%
\thanks{This work was supported by Delta-NTU Corporate Lab for Cyber-Physical Systems with funding support from Delta Electronics Inc. and the National Research Foundation (NRF) Singapore under the Corp Lab@University Scheme.}}
\titlerunning{Challenges in Digital Twin Development for CPPS}
%
%
\author{Heejong Park\inst{1}\orcidID{0000-0001-8979-2283} \and
Arvind Easwaran\inst{1}\orcidID{0000-0002-9628-3847} \and
Sidharta Andalam\inst{2}\orcidID{0000-0001-6660-8172}}
\authorrunning{Park et al.}
%
\institute{Nanyang Technological University, 50 Nanyang Avenue 639798, Singapore \\
\email{\{hj.park,arvinde\}@ntu.edu.sg}\\ \and
Delta Electronics, 50 Nanyang Avenue 639798, 50 Nanyang Avenue 639798, Singapore \\
\email{sidharta.andalam@deltaww.com}}
\maketitle              
\begin{abstract}

The recent advancement of information and communication technology makes digitalisation of an entire manufacturing shop-floor possible where physical processes are tightly intertwined with their cyber counterparts. This led to an emergence of a concept of digital twin, which is a realistic virtual copy of a physical object. Digital twin will be the key technology in Cyber-Physical Production Systems (CPPS) and its market is expected to grow significantly in the coming years. Nevertheless, digital twin is still relatively a new concept that people have different perspectives on its requirements, capabilities, and limitations. To better understand an effect of digital twin's operations, mitigate complexity of capturing dynamics of physical phenomena, and improve analysis and predictability, it is important to have a development tool with a strong semantic foundation that can accurately model, simulate, and synthesise the digital twin. This paper reviews current state-of-art on tools and developments of digital twin in manufacturing and discusses potential design challenges.

\end{abstract}

\section{Introduction}

The advancement of today's information and communication technologies (ICT) has enabled a collection and effective use of big data which give useful insights about various industrial assets and their operations. High availability of low-cost, low-powered sensors and Internet-of-Thing (IoT) devices together with their communication networks are the key enablers of cyber-physical systems. Cyber-phycial system is the main technological concept of the fourth industrial revolution, so called Industry 4.0~\cite{kagermann2013recommendations}, characterised by a tight integration of computations in a cyber world with physical processes in a real world. In cyber-physical system, changes in a physical process affect computations in a cyber world or vice-versa~\cite{lee2008cyber} where ICT enables a feedback between these two. A concept of digital twin, which is a ultra-realistic virtual counterpart of a real-world object, was introduced firstly by Grieves in 2003~\cite{grieves2017digital}. Since the introduction of Industry 4.0 at the Hannover Fair in Germany in 2013, the capability the digital twin in cyber-physical system has been received a great attention along with the recent advancement information technologies. The digital twin market is likely to be worth USD 15.66 billion by 2023 at a compound annual growth rate of 37.87\%~\cite{website:mam-dt}. 

Implementation of cyber-physical system in industrial manufacturing is called Cyber-Physical Production System (CPPS). In this domain, the use of digital twin has been mainly studied to improve time-to-market and MRO (Maintenance, Repair and Overhaul) costs, predict potential failures, and estimate remaining life of individual components, through high-fidelity simulation as well as real-time monitoring and control of manufacturing process. Since digital twins are built using the best available ICTs to mirror the physics of target objects in a virtual world, they can perform various simulations as if physical systems are tested in a real-life situation. Furthermore, digital twins can pause, resume, save, and restore their states to validate various corner-cases, which would rather be time-consuming and costly, if not impossible, to accomplish with physical systems. It is particularly valuable for the organisations who cannot afford very expensive resources to conduct exhaustive testing that may ruin their physical prototype or result in catastrophic events. For example, NASA and U.S. proposed digital twin concept to accelerate development of their future vehicles~\cite{glaessgen2012digital}.

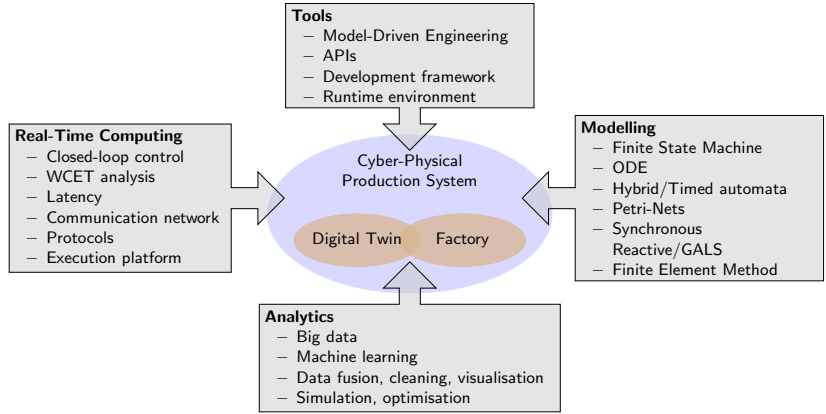
\begin{figure}[t!]
	\centering
	\sffamily
\scalebox{0.7}{
\begin{tikzpicture}[rotate border/.style={shape border rotate=#1},arrow box style/.style={draw,arrow box,arrow box shaft width=0.4cm,arrow box head extend=0.3cm,arrow box arrows={#1}}]
	\coordinate (cpps) at (0,0);
	\path (cpps) ++(-1,-0.5) coordinate (dt);
	\path (cpps) ++(1,-0.5) coordinate (factory);
	
	\draw[fill,color=blue!15] (cpps) ellipse [x radius=2.7cm,y radius=1.5cm];
	\draw[fill,color=brown!60,opacity=0.7] (dt) ellipse [x radius=1.2cm,y radius=0.5cm];
	\draw[fill,color=brown!60,opacity=0.7] (factory) ellipse [x radius=1.2cm,y radius=0.5cm];
	
	\node at (cpps)[above=10pt,align=center]{Cyber-Physical\\Production System};
	\node at (dt) {Digital Twin};
	\node at (factory) {Factory};
	\node at (-5.5,0.3)[arrow box style=east:1cm,text width=4cm,fill=gray!40,fill opacity=0.5, text opacity=1] {
		\textbf{Real-Time Computing}
		\begin{itemize}[topsep=0pt]
		\item Closed-loop control
		\item WCET analysis
		\item Latency
		\item Communication network
		\item Protocols
		\item Execution platform
		\end{itemize}
	};
	\node at (0,1.2)[arrow box style=south:0.8cm,fill=gray!40,fill opacity=0.5,text opacity=1,text width=4.5cm,anchor=south arrow tip] {
		\textbf{Tools}
		\begin{itemize}[topsep=0pt]
		\item Model-Driven Engineering
		\item APIs
		\item Development framework
		\item Runtime environment
		\end{itemize}
	};
	\node at (5.5,0.3)[arrow box style=west:1cm,rotate border=180,fill=gray!40,fill opacity=0.5,text opacity=1,text width=4.5cm]{
		\textbf{Modelling}
		\begin{itemize}[topsep=0pt]
		\item Finite State Machine
		\item ODE
		\item Hybrid/Timed automata
		\item Petri-Nets
		\item Synchronous Reactive/GALS
		\item Finite Element Method
		\end{itemize}
	};
	\node at (0,-0.9)[arrow box style=north:0.8cm,fill=gray!40,fill opacity=0.5,text opacity=1,text width=5.5cm, anchor=north arrow tip] {
		\textbf{Analytics}
		\begin{itemize}[topsep=0pt]
		\item Big data
		\item Machine learning
		\item Data fusion, cleaning, visualisation
		\item Simulation, optimisation
		\end{itemize}
	};

\end{tikzpicture}
}
	\caption{Components of Cyber-Physical Production System}
	\label{fig:bbcpps}
\end{figure}

An overview of technologies that create digital twin in CPPS is shown in Fig.~\ref{fig:bbcpps}. CPPS is a combination of both logical and physical components that can be characterised by continuous and discrete dynamics. In addition, modelling and implementing a digital twin may require skills from multiple disciplines, such as electromagnetism, fluid dynamics, and kinematics etc., to capture physical properties of the manufacturing process. Therefore, a modelling technique with varying levels of abstraction would be needed for both flexibility and expressiveness. The tight integration with the physical system often put real-time constraints on the operations of digital twin. As a result, designers would also need to consider real-time aspects of the twin such as worst-case execution time and communication latency in the time-analysable networks. Data measured from sensors and simulation are useful for predictive maintenance and optimising production process. A platform for data analytics that enriches digital twin capabilities is also a significant part of the CPPS. Lastly, a proper software framework will be required that incorporates APIs, runtime environment, and model-driven engineering.

Although digital twin is employed for tackling various problems~\cite{negri2017review}, it is still relatively a new research area where there exist several open research questions that have not yet been thoroughly explored such as:

\begin{enumerate}
    \item How to build a highly-accurate, yet scalable, digital twin for both simulation and real-time closed-loop control.
    \item How to mitigate issues related to \textit{uncertainties} and \textit{discrepancies} between the twin and the physical plant.
    \item How digital twin merges with big data. Are big data related technologies such as machine learning and statistical approaches part of digital twin or separate services? What are use-case scenarios of digital twin with big data?
    \item How to evaluate a digital twin that it faithfully mirrors its twinned system, how to quantitatively measure and compare two or more different digital twins of the same factory?
    \item What are the requirements of a digital twin development tool that addresses the aforementioned questions.

\end{enumerate}

The purpose of this paper is to review current state-of-art on tools and developments of digital twin in CPPS, discuss open research problems and suggest potential directions to address the aforementioned gaps.


The rest of this paper is organised as follows. Section~\ref{sec:dt-in-literature} presents literature review on digital twin architectures and modelling strategies. Section~\ref{sec:dt-sm} discusses an overview of digital twin in the context of smart manufacturing and CPPS. Section~\ref{sec:moc} introduces different models of computation. General-purpose as well as field-specific modelling tools are presented in Section~\ref{sec:tools}. An industrial manufacturing case study with open research questions are presented in Section~\ref{sec:case}. Finally, conclusions are given in Section~\ref{sec:conclusions}.






\section{Literature Review}\label{sec:dt-in-literature}

\begin{table}[t!]
    \centering
    \scriptsize
    \caption{Digital twin found in literature}
    \label{tab:literature}
    \rowcolors{1}{}{gray!20}
    \setlength\aboverulesep{0.5pt}
    \setlength\belowrulesep{0.5pt}
    \scalebox{0.87}{
    \begin{tabular}{ccv{2cm}v{5.7cm}cv{3cm}}
	\toprule
    \textbf{Ref} & \textbf{Year} & \textbf{Domain} & \textbf{Purpose} & \textbf{Big data} & \textbf{Tools used} \\
    \midrule
    \cite{larsen2016integrated} & 2016 & General CPS & Model-based design tool for multidisciplinary, collaborating modelling of CPS  & No & Modelio, Overture, Modelica, 20-sim, FMI\\
    \cite{tao2017digital} & 2017 & Manufacturing & Predictive maintenance, simulation, resiliency & Yes & \NA \\
	\cite{alam2017c2ps} & 2017 & Vehicle telematics & Optimising communication cost & No & Qfsm \\
	\cite{lopez2017software} & 2017 & Manufacturing & Predictive maintenance, application development, simulation.  & Yes & \NA \\
	\cite{ansys-2017-dt-pump} & 2017 & Process plants & Predictive maintenance, monitoring, 3D visualisation, simulation & Yes & ANSYS Simplorer, SCADE, PTC ThingWorx \\
	\cite{bliudze2017rigorous} & 2017& General CPS & CPS implementation & No & \NA  \\
	\cite{li2017dynamic} & 2017 & Structural health monitoring & Predictive maintenance & No & \NA \\
	\cite{schluse2018experimentable}& 2018 & Manufacturing & Simulation-based systems engineering & No & SysML \\
	\cite{tao2018digital} & 2018 & Manufacturing & Product lifecylcle management & Yes & \NA \\
 	\cite{potok2018sdcworks} &2018 & Manufacturing & Modelling, analysis, simulation & No & \NA  \\
 	\cite{scaglioni2018towards} &2018 &Manufacturing & Modelling, simulation & No & Modelica, FEM \\
    \bottomrule
    \end{tabular}
	}
\end{table}

Digital twin is the cornerstone of the Industry 4.0 wave which has been an active research area in the past several years. Table~\ref{tab:literature} shows some of the recent literature where digital twin is introduced for addressing various problems in a number of domains. The use of different modelling tools and whether the author highlights the use of big data analytics in their work are also indicated in the table.



An architecture of digital twin shop-floor (DTS) is presented in~\cite{tao2017digital}. There are four main components in this architecture: physical shop-floor (PS), virtual shop-floor (VS), shop-floor service system (SSS), and shop-floor digital twin data (SDTD). VS is a digital twin of the PS and data generated from both PS and VS are merged into the SDTD database. SSS consists of many sub-services, which are transformed into composite services based on demands from PS and VS. VS is used to simulate, predict, and perform calibration using the real-time data generated from PS. While the work gives a good overview of digital twin based physical and virtual space interconnection, authors do not tackle modelling VS directly although they suggest a number of tools that can be used to model VS in different levels of hierarchy: geometry, physics, behaviour, and rule. The same research group also showed employing digital twin for product development and for managing entire product life-cycle~\cite{tao2018digital,qi2018digital}. In particular,~\cite{qi2018digital} presented similarities and differences between digital twin and big data technologies and how they can be complementary with each other to enhance an overall manufacturing process.

A reference model for digital twin architecture for the cloud-based cyber-physical systems is presented in~\cite{alam2017c2ps}. The architecture consists of three intermediary layers, namely cyber-things layer, peer-to-peer communication layer, and intelligent service layer. In this work, closely related physical and cyber \textit{things} can create communication groups and peer-to-peer communication channels between those things are formed based on networking or communication criteria using a Bayesian belief network.

In~\cite{schluse2018experimentable}, authors propose a concept of experimentable digital twin (EDT) that combines the ideas of digital twins with model-based systems engineering and simulation technology. In addition to digital twin itself, EDT also comprises of a model of external environment that interacts with the twin via simulated sensors and actuators. The authors showed modelling of automotive headlight housing assembly using the EDT approach.

A case study for modelling an industrial machine tool is presented in~\cite{scaglioni2018towards}. In this work, authors use Finite Element Method (FEM) based preprocessing approach for modelling structural flexibility of machine's components. The models of cutting process, kinematic chains, and control systems are developed using Modelica~\cite{fritzson2010principles}.

An integrated tool-chain for model-based design of cyber-physical systems is introduced in~\cite{larsen2016integrated}. The tool enables co-simulation of multi-domain models by providing an integrated framework that combines multiple tools. They use Unifying Theories of Programming (UTP)~\cite{hoare1998unifying} as a foundation to give semantics to their heterogeneous approach.

ANSYS and PTC worked together to demonstrate how a digital twin of a pump can help diagnose and solve operating problems faster. The pump model is developed using ANSYS's Simplorer and SCADE whereas PTC's ThingWorx platform, which provides data collection and analysis services, is used to create an IoT ecosystem for devices and sensors~\cite{ansys-2017-dt-pump}.

A concept of software-defined control (SDC) has emerged recently~\cite{lopez2017software,potok2018sdcworks} which is inspired from the traditional software-defined networking (SDN), for programmatically configuring a communication network. A basic idea of SDC is to provide a central controller that has a global view of a system and separate decision making logic from the operations management solutions. In this framework, digital twin is suggested as a core simulation engine to improve decision making and detect faults in the manufacturing systems.

A concept of dynamic Bayesian network based digital twin is introduced in in~\cite{li2017dynamic} for monitoring the health of an aircraft. The authors propose a modification to the traditional probabilistic network that significantly reduces the computational cost for Bayesian interference. The approach integrates physics models with sources of uncertainty to predict crack growth on the airframe.

Authors in~\cite{bliudze2017rigorous} discuss modelling, discretisation, executability, simulation, and implementation of cyber-physical systems. The paper highlights the need of methods and tools with appropriate design languages underpinned by a solid semantic foundation which can model complex electromechanical systems. Various modelling techniques and respective challenges are discussed including detection of Zeno behaviours, difficulties in simulation of hybrid models in the presence of differential algebraic equations (DAE), physical systems modelling using linear and bond graphs. Although the authors do not particularly relate their work with digital twin development, the techniques and the design flow presented in the paper should definitely be considered in any digital twin development tools.

In literature, it is shown that digital twin is appeared the most in the manufacturing domain for predictive maintenance and simulation purposes. Several works~\cite{tao2017digital,tao2018digital,lopez2017software} propose the digital twin based software architectures and foresee the importance of big data analysis and its linkage with the digital twin. The works in~\cite{bliudze2017rigorous,li2017dynamic,scaglioni2018towards,potok2018sdcworks,larsen2016integrated} focus more on the modelling aspects. Nevertheless, implementation details of the software architectures and their use case scenarios are somewhat abstracted. In addition, most of works lack a comprehensive modelling framework except~\cite{potok2018sdcworks}, whose target is yet limited to discrete event systems, and~\cite{larsen2016integrated}, which focuses more on simulation of cyber-physical systems in general rather than digital twin itself. The next section discusses digital twin in the context of CPPS.




\section{Digital Twin in Manufacturing}\label{sec:dt-sm}

\begin{figure}[t!]
    \centering
    \includegraphics[scale=0.508,page=1]{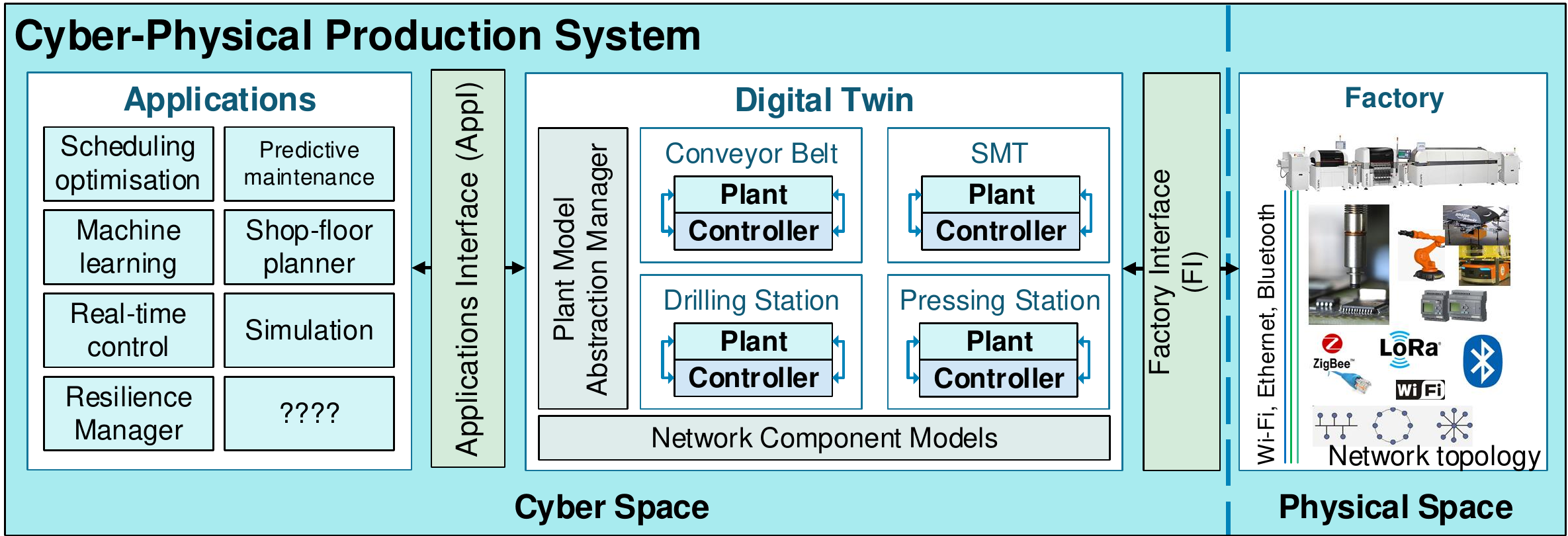}
    \caption{An overview of the proposed Cyber-Physical Production System (CCPS) architecture}
    \label{fig:cps}
\end{figure}

In literature and industries, the terms ``Smart Manufacturing'' and ``Industry 4.0'' are being used interchangeably and they are now almost synonymous with each other. The main objective is to leverage the recent advancement of information and communication technologies, such as cloud computing, IoT, and big data, to achieve autonomous, self-optimising, and self-diagnosing capabilities that can mitigate various problems in complex manufacturing scenarios. Digital twin is typically used in the context of cyber-physical systems to mirror the life of its real object via the best available physical models and sensor data. As a result, the digital twin enables simulation of real-world scenarios in a cyber world that otherwise would cost considerable amount of resource and time. Fig.~\ref{fig:cps} shows our proposed digital twin architecture for CPPS.


CPPS is a mechanism used in smart manufacturing and Industry 4.0 design principles. It is comprised of five main components: (1) a factory, (2) a digital twin and its runtime environment, (3) a factory interface to extract sensor/actuator data from the physical space, (4) an application interface that provides Application Programming Interfaces (APIs) to applications that wish to utilise the digital twin, and (5) the application themselves.

Most often a factory, also called a \textit{physical plant}, is a hybrid system, which is characterised with both continuous and discrete dynamics, modelled with a system of ODEs and DAEs, and state transitions, respectively. Examples of continuous dynamics of the plant are a movement of a workpiece on a conveyor belt and a movement of mechatronic arms during an assembly operation. On the other hand, discrete dynamics such as switching the operation mode from active to idle when no workpiece is detected for a certain amount of time.

The Plant Model Abstraction Manager and the Network Component Models are the parts of the digital twin runtime environment which manage lifecycles of individual twins and communication channels between them. Each twin consists of models of a plant and a controller that form a closed-loop control system via feedback and control signals. Interactions between these two models in the cyber space will also be accurately reflected in the physical space through the Factory Interface (FI).

The framework also provides a set of APIs that allow user applications to interact with the digital twin via the Application Interface (AppI). In this paper, we do not focus on the design of applications that utilise the digital twin, but rather on the digital twin development itself. However, it is worth to note here that technologies such as big data analysis can be employed in the application to deduce useful insights about the physical plant. The digital twin can communicate with this application to enhance its functionalities, for example, predicting potential failures and optimising throughput by adjusting its parameters. Similarly, the quality of data analysis can also be improved through data fusion of the physical plant and digital twin simulation.

Digital twin can be modelled using various levels of abstraction. Depending on requirements and resource availability, designers may choose a low-fidelity model such as finite-state machine or a higher-fidelity hybrid automaton or finite element method (FEM). Specification of plant components at different levels fidelity allows generation of \textit{mixed-fidelity digital twin}, which is a trade-off between accuracy and scalability. This trade-off makes implementing CPPS in bigger scale more practical since it would be computationally expensive to realise every aspect of physical plants using high-fidelity modelling approaches. 


\begin{table}[t!]
    \centering
    \setlist[itemize]{noitemsep,topsep=0pt,after=\vspace{-\baselineskip}}
    \rowcolors{1}{}{gray!20}
    \caption{Digital twin use case scenarios}
    \scriptsize
    \label{tab:dt-usecase}
    \scalebox{0.9}{
    \begin{tabular}{m{4cm}m{9cm}}
    \toprule
        \bfseries Features & \bfseries References \\ 
        \midrule
        Root-cause failure analysis and predictive maintenance & \begin{itemize}
            \item Detection of a faulty valve~\cite{ansys-2017-dt-pump}.
            \item Data-oriented analysis and prediction for wind turbines~\cite{ge-wind-farm}.
        \end{itemize}\\
        High-fidelity simulation & 
        \begin{itemize}
            \item Water pump simulation~\cite{ansys-2017-dt-pump}.
            \item Financial and risk simulation~\cite{ge-wind-farm}.
            \item Testing machineries for filling and packaging medications~\cite{siemens-bausch}.
            \item Simulation of a sheet metal punching machine~\cite{moreno2017virtualisation}.
        \end{itemize}\\
        Closed-loop real-time control & 
        \begin{itemize}
            \item Real-time control of the water pump~\cite{ansys-2017-dt-pump}.
            \item Turbine control~\cite{ge-wind-farm}.
            \item Human-robot collaborative assembly system~\cite{wang2017human}.
        \end{itemize}\\
        3D visualisation & 
        \begin{itemize}
            \item 3D simulation model that shows cavitation inside a waterpump~\cite{ansys-2017-dt-pump}.
            \item 3D visualisation of pharmaceutical machines~\cite{siemens-bausch}.
            \item 3D model of a brake pad wear~\cite{magargle2017simulation}.
            \item 3D visualisation of the punching machine process~\cite{moreno2017virtualisation}.
        \end{itemize}\\
    \bottomrule
    \end{tabular}
	}
\end{table}

In manufacturing, employing a digital twin based cyber-physical system is favourable in a variety of scenarios. Table~\ref{tab:dt-usecase} shows examples of digital twin use case scenarios, which are grouped in four different categories. Since digital twins are often developed to handle multiple problems, most of the works overlap with each other in those categories: 

\begin{enumerate}
    \item \textit{Root-cause failure analysis and predictive maintenance}: When there is a fault in the system, operators can utilise digital twin to pinpoint a root-cause thanks to its rich structural and behavioural information about the twinned system. In this case, choosing an appropriate formal \textit{model of computation} (MoC) of the plant would play a pivotal role that facilitates semantic preservation between different design phases. From this, the digital twin will be able to back-trace the failure from an executable code all the way back to the modelling phase. 
    
    \item \textit{High-fidelity simulation}: Data analytics, storage, and together with high-fidelity modelling techniques will enable the digital twin to assess different manufacturing as well as fault scenarios which provides highly accurate results that would be difficult to obtain via traditional simulation techniques. 
    
    \item \textit{Closed-loop real-time control}: Digital twin augments the quality of the classical plant and controller closed-loop control where it can add additional values such as dynamicity, reconfigurability, connectivity, global intelligence, and predictability. This is possible because the digital twin can directly influence a behaviour of the physical plant and also acts as an intermediary entity between the plant and the user applications which consume the twin as a service. In this context, we foresee that the digital twin has a great chance to become a technological gateway to a wide range of cyber-physical system applications. 
    \item \textit{3D visualisation}: The multi-domain and multi-physics design approaches make it natural for the digital twin to be a good candidate for 3D visualisation for interactive validation and inspection. 
\end{enumerate}



In the next section, a number of well-known modelling strategies is presented that could be a back-end of the digital twin development framework.

\begin{figure}[t!]
    \centering
    \begin{subfigure}{0.5\textwidth}
    \setlength\abovecaptionskip{0.15cm}
    \scriptsize
    \def\arraystretch{1.3}
    \checkmark: Supported \halfcheck: Supported via extension \cross: Not supported
    
    \scalebox{0.9}{
    \begin{tabular}{c|c|c|c|c|c|c}
        \toprule
        \multirow{2}{*}{MoCs}& \multirow{2}{*}{\makecell{Physical\\time}} & \multirow{2}{*}{Hierarchy} & \multicolumn{2}{c|}{Concurrency} & \multirow{2}{*}{Data} & \multirow{2}{*}{ODEs} \\
        \cline{4-5}
        &&& Sync & Async && \\
        \hline
        FSM & \cross & \cross & \cross & \cross &\halfcheck & \cross\\
        PN & \cross & \checkmark & \cross & \checkmark &\cross &\cross\\
        CPN & \cross & \checkmark & \cross &\checkmark &\checkmark &\cross \\
        TPN & \checkmark & \checkmark & \cross & \checkmark & \cross & \cross\\
        SR & \halfcheck & \checkmark & \checkmark & \cross & \checkmark & \cross \\
        GALS & \halfcheck & \checkmark & \checkmark & \checkmark & \checkmark & \cross \\
        TA & \checkmark & \cross & \halfcheck & \cross & \halfcheck & \cross \\
        HA & \checkmark & \cross & \cross & \cross & \halfcheck & \checkmark \\
        \bottomrule
    \end{tabular}%
    }
	\vspace{13pt}
    \caption{A feature comparison among different MoCs}
    \label{fig:mocs}
    \end{subfigure}\hfill%
    \begin{subfigure}{0.47\textwidth}
    \pgfdeclarelayer{bg}    
    \pgfsetlayers{bg,main} 
    \contourlength{1.2pt}
    \scalebox{0.75}{
    \begin{tikzpicture}[every node/.style={color=black}]
        \begin{axis}[
            xlabel=$Observability$,
            ylabel=$Fidelity$,
            xmax=10,ymax=10,
            xmin=0,ymin=0,
            ytick distance=1, xtick distance=1,
            axis x line=bottom, axis y line=left,
            ticks=none,
        ]
        \draw[->,color=gray] (5,1) node[below] {} -- (5, 9) node[above] {Accurate};
        \draw[<->,color=gray] (1,5) node[below right=0 and -15pt,align=center] {Uncertain} node[above right=0 and -17pt] {Open-loop} -- (9, 5) node[below left=0 and -20pt] {Highly-certain} node [above left=0 and -19pt] {Closed-loop};
        \draw (5,5) node[opacity=1] {\sffamily \contour{green!20}{PN,CPN,SR,GALS}};
        \begin{pgfonlayer}{bg}
            \fill[blue!10] (5,2.5) circle[x radius=3.3cm, y radius=0.4cm] 
                (1.5,2.5) node {\sffamily FSM};
            \fill[green!80,opacity=0.3] (5,5) circle[x radius=3.3cm, y radius=0.7cm];
                
            \fill[red!20] (5,7) circle[x radius=3.3cm, y radius=0.4cm]
            	(2.5,7) node {\sffamily TA,TPN};
            \fill[orange!20,opacity=0.8] (5,8) circle[x radius=3.3cm, y radius=0.2cm] 
                (2.5,8) node {\sffamily HA};
        \end{pgfonlayer}
        \end{axis}
    \end{tikzpicture}%
    }
    \caption{Relationship between model fidelity and observability of a physical plant}
    \label{fig:fidel}
    \end{subfigure}
    \caption{Models of computation and observability of a production system}
    \label{fig:fidelities}
\end{figure}
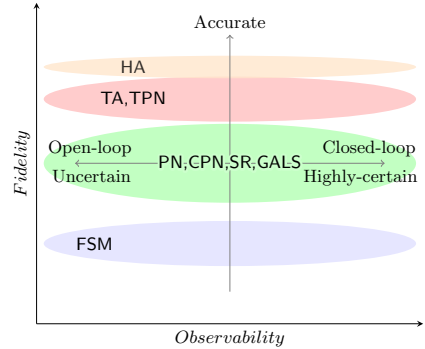

\section{An Overview of Models of Computation}\label{sec:moc}

Model of computation (MoC) deals with a set of theoretical choices that build an execution model of the design language. MoCs define how computations are carried out, for example sequentially or concurrently, interaction/synchronisation between computational units, and a notion of time, etc., without binding them to specific implementation details. Programming language equipped with formal MoC ensures that the resulting program follows the semantics of the corresponding MoC. On the other hand, a programming model for a certain MoC can also be built on a top of host programming language, for example through a library or a framework.

Many formal MoCs available for modelling a digital twin~\cite{brooks2008heterogeneous}. Some of them are shown in Fig.~\ref{fig:fidelities}, which can be basis of the digital twin development framework. Each MoC supports different features: \textit{Physical time} determines an ability for a MoC to relate its computations with a continuously evolving quantity, typically in real or integer domains. A simple example is when a Timed Automata (TA)~\cite{alur1991theory} generate an output event after a clock value reaches 5. \textit{Hierarchical composition} describes an ability to compose two or more basic design entities to achieve a more advanced and complex functionality. This can be easily found in many digital systems designs, for example an arithmetic logic unit composed of adders and subtractors, etc. \textit{Concurrency} describes an ability of basic entities in a model that can be executed in overlapping time. Typically, it refers to either synchronous or asynchronous concurrency. \textit{Data} indicates whether a MoC supports for variables, expressions, constructs, type systems, and etc., to perform data-oriented algorithmic computations. Lastly, \textit{ODEs} refer to an ability to capture continuous dynamics of a system via a system of ordinary differential equations. There is no MoC that supports all the features. An FSM can be used to capture various control-dominated behaviours. However, it alone is not well suitable for accurately capturing a complex nature of physical systems that are inherently concurrent, time-dependent, and data-rich. 

Petri-Nets~\cite{peterson1981petri} are good models for describing concurrent and distributed systems. However, it lacks expressiveness in data computations and timing properties. Coloured Petri-Nets (CPN)~\cite{jensen2015colored} and Timed Petri-Net (TPN)~\cite{ramchandani1974analysis} are extensions of the PN with data and timing features, respectively. Nevertheless, they lack in capturing continuous dynamics of a system. Although the original formal definition of the PN do not include hierarchy, there is an extension of PN that supports hierarchical composition~\cite{fehling1991concept}.

Synchronous Reactive (SR) MoC~\cite{berry1992esterel} provides a set of constructs for capturing reactive behaviours of a system. The execution semantics of SR constructs underpinned with rigorous mathematical foundation enables bug avoidance in the early design phase using correct-by-construction compilation and verification techniques~\cite{berry1992esterel}, which increase confidence in the correctness of final designs. While SR MoC has a notion of time, they are only logical. Moreover, the SR MoC is not amenable to a large distributed systems since every design component is synchronised with a single global clock. 

Globally Asynchronous Locally Synchronous (GALS)~\cite{malik2010systemj} is a superset of the SR MoC where a system is modelled using several synchronous subsystems, which run asynchronously and communicate with each other via a message passing mechanism. Still, the most of GALS-based modelling languages often lack an ability to model real-time behaviours and continuous dynamics of many physical systems. 

Hybrid Automata (HA)~\cite{henzinger2000theory} supports modelling of both discrete and continuous dynamics of a system. It uses a state machine with a finite set of real-valued variables whose values evolve according to a set of ODEs. A notion of reference time can also be defined in HA using a simple formula such as $\dot{t}=1$. However, it is a flat structure similar to an FSM, and does not directly support concurrency and hierarchical composition.

Generating a highly accurate, time-synchronised, and scalable digital twin, therefore requires a technique that combines various MoCs to complement each other's weaknesses. Generally, fidelity of a model is based on how well it can capture dynamics of physical phenomena. On the other hand, effectiveness of a modelling technique varies depending on \textit{observability} of the physical plant. As shown in Fig.~\ref{fig:fidel}, the relationship between fidelity of a model and observability of a physical plant categorises the digital twin based CPPS into largely four groups:

\begin{enumerate}
    \item \textit{Accurate but uncertain model}: High-fidelity models can accurately capture a plant's behaviour according to the specification. However, since a less observable plant only provides feedback at certain discrete instants, there is a low confidence on dynamics of the plant between two feedback events. For example, a digital twin can only speculate the current position of an item that moves along a conveyor belt based on the motor speed. Any unexpected events that occur between two photoelectric sensors cannot be captured by the digital twin unless the item is enabled with RFID tracking.
    \item \textit{Less accurate and uncertain model}: In this case, a model does not capture the plant's behaviour accurately and the plant is also less observable. An example of this scenario is using an untimed model such as  FSM for modelling a change of the temperature of the boiling water and the plant only provides feedback when the temperature reaches at $T_{max}$ threshold.
    \item \textit{Less accurate and highly-certain model}: In this scenario, the plant is fully observable where it provides changes its states in a frequent manner. Low-fidelity models, however may not fully utilise such information. For example, for a multi-axis arm movements, discrete event models may only react to events when the arm only reaches its final destination of the movement, ignoring how it reaches there.
    \item \textit{Accurate and highly-certain model}: This is when both the model and the plant fully synchronises and the model can accurately trace dynamics of the plant at any instants of time.
\end{enumerate}

%


In the next section, an overview of existing modelling tools are presented. These tools are used for general cyber-physical system development, but also can target digital twin.

\section{Modelling Tools}\label{sec:tools}

\mbox{Table}~\ref{tab:tools} summarises various types of modelling tools available which target general-purpose discrete and continuous systems as well as more application specific fields. It is not possible to cover all of them here, but we selected some of the notable tools currently available in the market and academia.

{
	\def\checkmark{\tikz\fill[scale=0.4](0,.35) -- (.25,0) -- (1,.7) -- (.25,.15) -- cycle;}
	\def\halfcheck{\tikz\draw[scale=0.4](0,0) -- (0.5,0) -- (0.5,0.5) -- (0,.5) -- cycle;}
	\rowcolors{1}{}{gray!20}
	
	\begin{table}[t!]
		\caption{A summary of modelling tools}
		\label{tab:tools}
		\setlength{\tabcolsep}{4pt}
		\scriptsize
		\renewcommand{\arraystretch}{0.6}
		\scalebox{0.9}{
			\begin{tabular}[]{>{\rfont}m{3cm}>{\rfont}m{3.5cm}>{\rfont}m{2.9cm}>{\rfont}m{2.5cm}}
				\toprule
				\hiderowcolors
				\setrowfont\bfseries Tools & Licence & Formal MoC & \shortstack[l]{Code\\ generation} \\
				\midrule
				\showrowcolors
				\setrowfont{} SCADE Simplorer\tablefootnote{https://www.ansys.com/products/systems/ansys-twin-builder}	& Commercial & Multi-paradigm  & --  \\
				SCADE Suite\tablefootnote{http://www.esterel-technologies.com/products/scade-suite/} & Commercial & Synchronous Dataflow  & \checkmark   \\ 
				ANSYS Twin Builder\tablefootnote{https://www.ansys.com/products/systems/ansys-twin-builder} & Commercial & Multi-paradigm & -- \\ 
				SCILAB\tablefootnote{https://www.scilab.org/} & GPLv2 & --  & via	thrid-party plugins: X2C, Project-P  \\
				Modelica families\tablefootnote{https://www.modelica.org/} & Open source and commercial & --  & \checkmark (some implementations)  \\
				MATLAB/Simulink\tablefootnote{https://www.mathworks.com/products/matlab.html} & Commercial & --  & \checkmark  \\
				BCVTB\tablefootnote{https://simulationresearch.lbl.gov/bcvtb} & Modified BSD & Actor-based  & --  \\
				INTO-CPS\tablefootnote{https://into-cps.github.io/} & Open source & VDM, SysML & \checkmark  \\ 
				CIF\tablefootnote{http://cif.se.wtb.tue.nl/index.html} & MIT & HA  & \checkmark  \\
				Flow*\tablefootnote{https://flowstar.org} & GPL  & HA  & --  \\
				Flexsim\tablefootnote{https://www.flexsim.com/} & Commercial & -- & -- \\
				HyST\tablefootnote{http://www.verivital.com/hyst/} & LGPL & HA &	 \checkmark (To other HA models)  \\
				SL2SX\tablefootnote{https://github.com/nikos-kekatos/SL2SX} & GPLv3 & --  & \checkmark (Simulink  to SpaceEx) \\
				SpaceEx\tablefootnote{http://spaceex.imag.fr/} & GPLv3 & HA  & --  \\
				IOPT\tablefootnote{http://gres.uninova.pt/IOPT-Tools/login.php} & Free (Web-based) & Petri-Net &  \checkmark  \\
				CPN Tools\tablefootnote{http://cpntools.org/} & GPLv2 & Coloured Petri-Net &  
				--  \\
				Ptolemy II~\tablefootnote{https://ptolemy.berkeley.edu} & Mixed & Actor-oriented on top of heterogenous MoC & \checkmark \\  
				Esterel\tablefootnote{http://www-sop.inria.fr/esterel-org/files/Html/News/News.htm}	& Mixed and commercial & Synchronous Reactive &  \checkmark  \\
				Lustre\tablefootnote{http://www-verimag.imag.fr/Sync-Tools.html?lang=en} & Free and commercial & Synchronous Dataflow  &	\checkmark  \\ 
				C\'{e}u\tablefootnote{http://www.ceu-lang.org/} & MIT & Synchronous Reactive & \checkmark \\
				\bottomrule
			\end{tabular}
		}
	\end{table}
}

Tools like Scilab, Modelica, and Matlab/Simulink provide general-purpose, numerical computing environment for modelling a wide range of systems such as mechanical and electrical systems, fluid dynamics, and etc. The SCADE Suite is a model-based development environment for mission critical embedded software with Lustre as its core language. ANSYS Simplorer and Digital Twin Builder provide multi-domain, co-simulation environment with support for VHDL-AMS, Modelica, C/C++, and SPICE languages along with MIL (Model-in-the-Loop), and SIL (Software-in-the-Loop) capabilities. The model can be connected to various industrial IoT platforms such as PTC ThingWorx, GE Predix, and SAP Leonardo.

There are also tools equipped with formal models of computation. For example, synchronous reactive languages such as Esterel, Lustre, and C\'{e}u and Petri-Net based IOPT and CPN Tools. Building Controls Virtual Test Bed (BCVTB) is a software environment based on Ptolemy II project~\cite{brooks2008heterogeneous}, to couple different simulation programs for co-simulation.

Modelling tools which target hybrid systems have direct support for capturing both discrete and continuous dynamics and transition between these modes. For example, Compositional Interchange Format (CIF) is a automata-based language for the specification of discrete event, timed, and hybrid systems. HyST and SL2SX do source-to-source transformation to enable evaluation of HA models using different tools. Model checking tools for HA models also exist, for instance SpaceEx and Flow*.

Flexsim is a simulation software for manufacturing factories including 3D visualisation and statistical reporting and analysis features. The Integrated Tool Chain for Model-based Design of Cyber-Physical Systems (INTO-CPS) is an integrated tool chain for comprehensive model-based design of cyber-physical systems. It aims to provide a framework for model-based design and analysis by combining multiple models generated from different tools using the Functional Mock-up Interface (FMI) standard~\cite{blochwitz2011functional}.

In the next section, a case study called the IMPACT manufacturing line is introduced where an initial concept of the digital twin modelling is presented.


\begin{figure}[b!]
	\centering
	\includegraphics[scale=0.30]{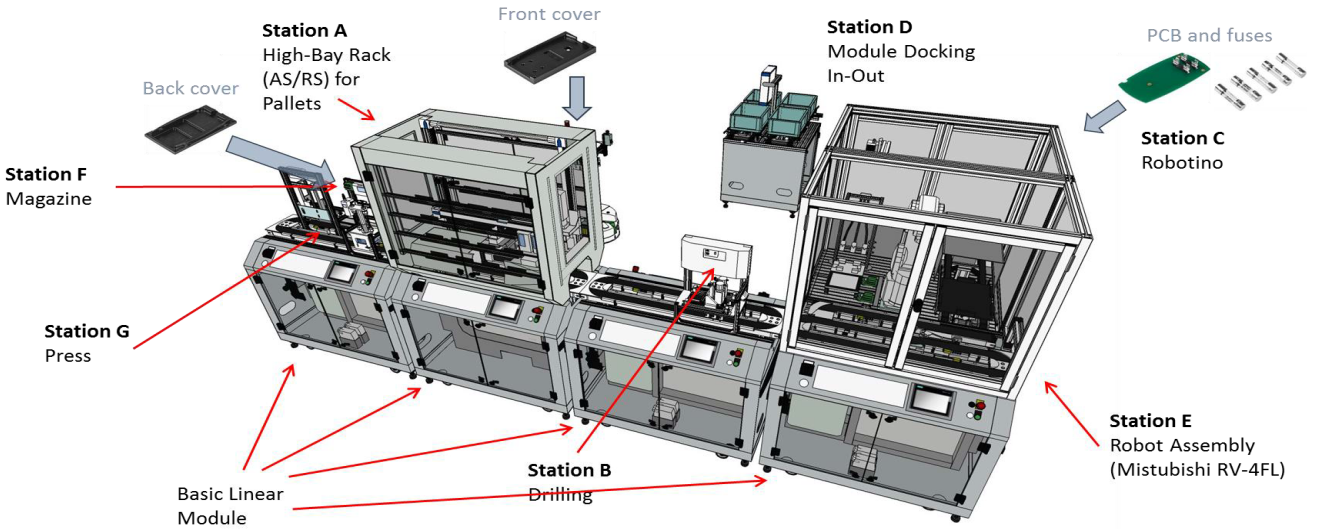}
	\caption{The IMPACT line -- A testbed for future manufacturing}
	\label{fig:impact}
\end{figure}

\section{Case Study: the IMPACT Manufacturing Line}\label{sec:case}

To illustrate the use of heterogeneous models and discuss open issues, we propose developing a cyber-physical production system case study called the IMPACT line shown in Fig.~\ref{fig:impact}. It consists of four linear modules with parallel conveyors and seven processing stations for manufacturing smart phones. The following outlines the operation of each station:

\begin{enumerate}
	\item \textit{High-Bay Rack} -- When there is a request from an
		external source for manufacturing a product, a cartesian robot is
		triggered to pick-up a workpiece pallet. The pallet is then placed
		on the conveyor belt. A motor for the conveyor belt is turned on to
		move the pallet to the drilling module
	\item \textit{Drilling Module} -- Two drilling spindles are advanced
		in the Z and X directions to make two pairs of holes into the
		workpiece.
	\item \textit{Robot Assembly Module} -- The 6-axis arm picks up the
		rear panel from the pallet and place it on the processing bay. The
		arm places a PCB on the panel and switches to a smaller gripper so
		that it can pick up and install a fuse into the PCB. The arm
		switches to the original gripper to place the rear panel with the
		finished PCB back to the pallet.
	\item \textit{Mobile Station} -- This module delivers
		boxes of PCBs to the IMPACT line.
	\item \textit{Robotino} -- This automated guided vehicle transfers a
		box of PCBs from the Mobile Station to the Robot Assembly Module.
	\item \textit{Magazine Module} -- This module places a front panel on
		the PCB.
	\item \textit{Pressing Module} -- This module applies pressing force
		to seal the product.
\end{enumerate}

This CPPS has a number of characteristics for demonstrating the need for digital twin development: (1) modelling continuous dynamics such as the 6-axis arm and cartesian robot movements, (2) choosing the right modelling strategies for the machines with different observabilities, (3) detecting faults such as unsatisfactory drilling operations due to wear and tear of a drill bit and misplacement of fuses on a PCB, (4) and real-time closed-loop control.

\subsection{Modelling Continuous Dynamics of the Factory}

\begin{figure}[b!]
	\centering
	\begin{tikzpicture}[every node/.style={align=left}, node distance=1cm,
	auto]
	
	\node[initial, state] (Idle) {$Idle$};
	\node[state] (On) [right=of Idle] {$On$};
	
	\path[->] (Idle) edge[bend left] node {$\frac{TurnOn}{v=0.03}$} (On)
	(On) edge[bend left] node{$\frac{\neg TurnOn\vee TimeOut}{v=0}$} (Idle)
	edge [loop right] node {$\frac{WP}{Reset}$} ()
	;
	\end{tikzpicture}
	\caption{An FSM model of the conveyor belt}
	\label{fig:cv-fsm}
\end{figure}
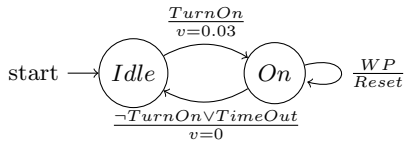

This section focuses modelling the motion of the conveyor belt on the linear modules and the cartesian robot arm in the High-Bay Rack for the illustration purpose. The first model of the conveyor belt is shown in Fig.~\ref{fig:cv-fsm}. This is the simplest possible case where the conveyor belt operates in either one of two macro states: $Idle$ or $On$ mode. When $TurnOn$ signal is set high by the controller, the machine makes transition to state $On$ after setting $v=0.03$, which indicates the speed of the conveyor belt. The conveyor belt stays on as long as an incoming workpiece $WP$ resets a timer via $Reset$. It goes $Idle$ mode when $TurnOn$ signal is unset by the controller or there were no incoming workpieces for the last $x$ time period indicated by $TimeOut$.


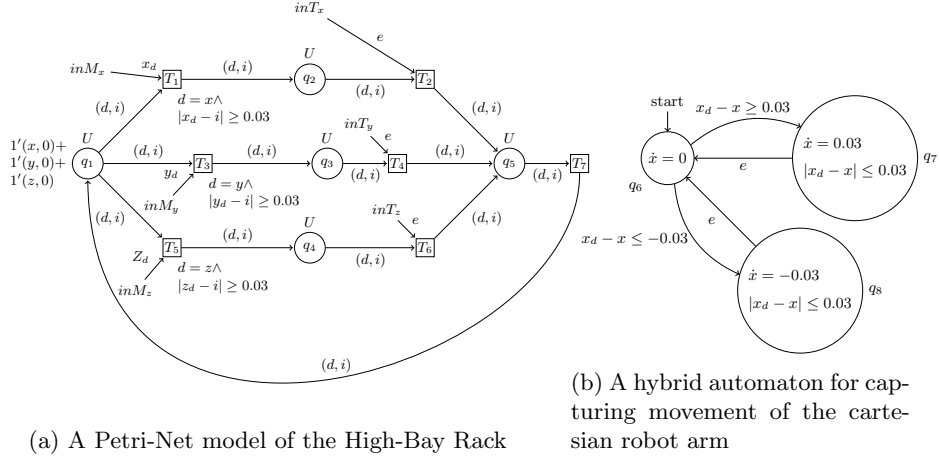
\begin{figure}[t!]
	\centering
	\begin{subfigure}[b]{0.6\textwidth}
	    \scalebox{0.6}{
		\begin{tikzpicture}[every node/.style={align=left}, node
			distance=2cm, auto]
			\clip (-2cm,4cm) rectangle (11.5cm,-5.5cm);
			\node[place,label=left:{$1'(x,0)$+\\$1'(y,0)$+\\$1'(z,0)$},label=above:$U$] (i) {$q_1$};
			\node[transition, above right=of i,label=273:{$d=x\wedge$\\$|x_d-i|\geq 0.03$}]
			(mx) {$T_1$} edge[pre] node[above left] {$(d,i)$}  (i);
			\node[place, label=above:$U$, right=2.5cm of mx] (W1) {$q_2$};
			\node[above=1.5cm of i] (input1) {$inM_x$} edge[post] node {$x_d$} (mx);
			\node[transition,right=of W1] (join1) {$T_2$} edge[pre] node {$(d,i)$} (W1);
			\node[place, below right=of join1, label=above:$U$] (joinp1) {$q_5$};
			\node[transition,right=1cm of joinp1] (fint) {$T_7$} edge[pre] node {$(d,i)$} (joinp1);

			\node[transition,right=of i, label=273:{$d=y\wedge$\\$|y_d-i|\geq 0.03$}]
			(my) {$T_3$} edge[pre] node[above]{$(d,i)$} (i);
			\node[below left=0.5cm and 0.2cm of my] (input2) {$inM_y$}
			edge[post] node {$y_d$} (my);
			\node[place, label=above:$U$, right=2.2cm of my] (W2) {$q_3$};

			\node[transition,below right=of i,label=273:{$d=z\wedge$\\$|z_d-i|\geq
			0.03$}] (mz) {$T_5$} edge[pre] node {$(d,i)$} (i);
			\node[below left=0.5cm and 0.1cm of mz] (input3) {$inM_z$} edge[post] node {$Z_d$}
			(mz);
			\node[place,label=above:$U$, right=2.5cm of mz] (W3) {$q_4$};

			\node[transition,right=1cm of W2] (join2) {$T_4$} edge[pre] node
			{$(d,i)$} (W2);
			\node[transition,right=of W3] (join3) {$T_6$} edge[pre] node {$(d,i)$} (W3);

			\node[above=1cm of W1] (input2) {$inT_x$} edge[post] node {$e$} (join1);
			\node[above left=0.4cm and 0.2cm of join2] (input3) {$inT_y$} edge[post] node {$e$} (join2);
			\node[above left=0.4cm and 0.2cm of join3] (input4) {$inT_z$} edge[post] node {$e$} (join3);

			\draw[->] (mx) edge[post] node {$(d,i)$} (W1)
								(my) edge[post] node {$(d,i)$} (W2)
								(mz) edge[post] node {$(d,i)$} (W3)
								(join1) edge[post] node {$(d,i)$} (joinp1)
								(join2) edge[post] node {$(d,i)$} (joinp1)
								(join3) edge[post] node[below right] {$(d,i)$} (joinp1)
								(fint) edge[post, bend left=90, in looseness=1.8]
								node[above] {$(d,i)$} (i);
		\end{tikzpicture}
		}
		\caption{A Petri-Net model of the High-Bay Rack}
		\label{fig:rbh-petri}
	\end{subfigure}\hfill%
	\begin{subfigure}[b]{0.37\textwidth}
	    \scalebox{0.65}{
		\begin{tikzpicture}[every node/.style={align=left}, node
			distance=2cm, auto]
			\node[initial, initial where=above, state,label=below left:$q_6$] (q6) {$\dot{x}=0$};
			\node[state] (q7)[right=of q6,label=right:$q_7$] {$\dot{x}=0.03$\\[5pt]$|x_d-x|\leq0.03$};
			\node[state] (q8)[below right=of q6,label=right:$q_8$] {$\dot{x}=-0.03$\\[5pt]$|x_d-x|\leq0.03$};

			\path[->] (q6) edge[bend left] node[above] {$x_d-x\geq0.03$} (q7)
										edge[bend right] node[left] {$x_d-x\leq-0.03$} (q8)
										 (q7) edge node {$e$} (q6)
										 (q8) edge node {$e$} (q6)
										 ;
		\end{tikzpicture}
		}
		\caption{A hybrid automaton for capturing movement of the cartesian
		robot arm}
		\label{fig:rbh-ha}
	\end{subfigure}
	\caption{Modelling the High-Bay Rack}
	\label{fig:rbh-model}
\end{figure}

The cartesian robot arm in the High-Bay Rack (HBR) is able to move in
x,y,z directions simultaneously. Let us assume a designer has chosen
Coloured Petri-Net (CPN) to model the robot since it features both
concurrency and data manipulation for capturing the position of the arm
in the cartesian space. The corresponding CPN model is shown in
Fig.~\ref{fig:rbh-petri}. This CPN has four colours (data types) and
two variables defined as follows.

\begin{lstlisting}
	type dir = | x | y | z
	type pos = int
	type U = dir * pos
	type E = e
	var d : dir
	var i : pos
\end{lstlisting}

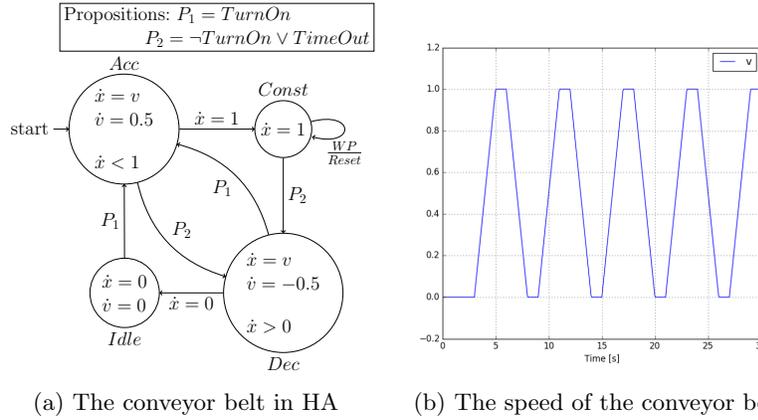
\begin{figure}[t!]
	\centering
	\begin{subfigure}[b]{0.4\textwidth}
		\scalebox{0.5}{
			\begin{tikzpicture}[every node/.style={align=left,font=\Large}, node distance=2cm,
			auto]
			
			\node[state, initial,label=above:$Acc$] (acc) {$\dot{x}=v$\\$\dot{v}=0.5$\\[5mm]$\dot{x}<1$};
			\node[state,label=above:$Const$] (const) [right=of acc] {$\dot{x}=1$}; 
			
			\node[state,label=below:$Dec$] (dec) [below=of const]
			{$\dot{x}=v$\\$\dot{v}=-0.5$\\[5mm]$\dot{x}>0$};
			\node[state] (idle) [label=below:$Idle$,left of=dec,node distance=120pt]
			{$\dot{x}=0$\\$\dot{v}=0$};
			
			\path[->] (acc) edge node (e_top){$\dot{x}=1$} (const)
			(const)	edge node {$P_2$} (dec)
			(dec) edge node  {$\dot{x}=0$} (idle)
			(idle) edge node {$P_1$} (acc)
			(acc) edge [bend right=30] node {$P_2$} (dec)
			(dec) edge [bend right=30] node {$P_1$} (acc)
			(const) edge[loop right] node [below=0.2cm] {$\frac{WP}{Reset}$} ();
			
			\node[above=1.5cm of e_top, draw] (legend) {Propositions:
				$P_1=TurnOn$\\\makebox[2.15cm]{}$P_2=\neg TurnOn\vee TimeOut$};
			\end{tikzpicture}
		}
		\caption{The conveyor belt in HA}
		\label{fig:conv-ha}
	\end{subfigure}
	\begin{subfigure}[b]{0.5\textwidth}
		\centering
		\includegraphics[trim=0.8cm 0cm 0.6cm 1cm,clip, scale=0.36]{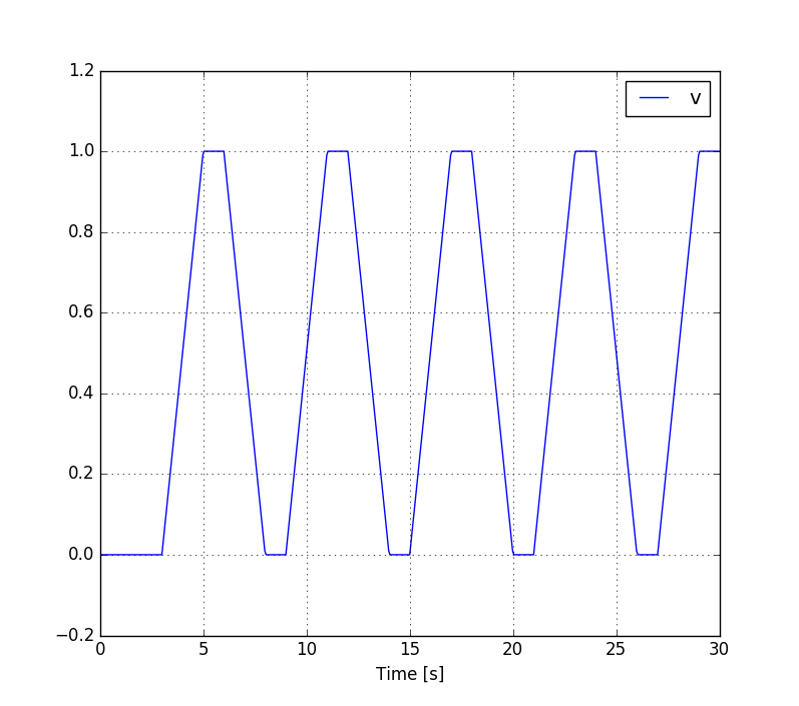}
		\caption{The speed of the conveyor belt}
		\label{fig:conv-graph}
	\end{subfigure}
	\caption{The conveyor belt refined}
	\label{fig:conv-refined}
\end{figure}

The three outgoing arcs from $q_1$ indicate the robot arm can move in
x,y,z directions simultaneously when it is requested via input signals
$inM_x$, $inM_y$, and $inM_z$. The initial marking at $q_1$ shows there
are three tokens of type $U$. The plus sign indicates the tokens are
combined into a multiset. Consider a token travelling from $q_1$ to
$q_5$ via transitions $T_1$ and $T_2$. The guard on $T_1$ ensures that
it is only enabled when there is a token of type $U$ in $q_1$ whose
first element is $x$. When there is a request to move the arm in x
direction, i.e. when $|x_d-i|\geq 0.03$, a token is added to the place $q_2$
indicating the model is now in the ``moving'' state. The colour set $U$
on $q_2$ specifies type of tokens which may reside on the place.

One of the requirements of the digital twin framework is to provide
a real-time view of the current state of the modelling plant. However at
this point, the designer realises that he/she cannot easily capture
continuous dynamics of the arm movement using CPN since it does not
support the flow of continuous variables. Assume that the tool allows to
mix different models so that the designer does not have to redesign the
robot arm from scratch. In this case, HA can be added that runs together
with the previously designed CPN as shown in Fig.~\ref{fig:rbh-ha}.
Here, the HA makes a transition from $q_6$ to $q_7$ or $q_8$ when the
final position $x$ is greater or smaller than the current coordinate,
respectively. States $q_6$ and $q_7$ show continuous evolution of
variable $x$ using its derivative $\dot{x}$, which is equivalent to $i$
of the token $(x,i)$ in $q_2$ in Fig.~\ref{fig:rbh-petri}. When the
arm reaches its destination, i.e. $|x_d-x|\leq0.03$, the HA makes
transition back to $q_6$ while generating an event $e$. This event
enables transition $T_2$ in Fig.~\ref{fig:rbh-petri} and results in
returning of the token back to $q_1$ for serving the next request for
$x$ direction movement.

By supporting different levels of abstraction, a model can be
enhanced further to incorporate finer details of operating modes. For
example, after turning on a conveyor from the  idle state, it may take
some time until the motor reaches full
rotational speed. Thus,	the original conveyor belt modelled in an FSM
can be further refined using HA, which is shown in
Fig.~\ref{fig:conv-ha}.  Here, the system starts with the state $Acc$
where the speed increases according to the flow variable $\dot{v}=0.5$.
When the speed reaches 1, a jump condition is satisfied that results in
a state transition to $Const$. The system stays in this state as long as
the next workpiece arrives before $TimeOut$ or a controller resets signal
$TurnOn$. The state $Dec$ decelerates the speed of the conveyor belt
until it reaches zero and the system stays in $Idle$ until $TurnOn$
signal is set by the controller. The change of speed of the conveyor
belt over the period of 30 seconds is shown in
Fig.~\ref{fig:conv-graph}, which is modelled in Modelica. Similar
approaches can be applied to model rest of the stations of the IMPACT
line. 




\subsection{Open Problems}

\subsubsection{Addressing Discrepancy in a Model.}

Discrepancy between a model and a physical system may arise due to several reasons. For example, increase in complexity, choosing an inadequate modelling strategy, accumulation of errors over time, lack of documentation about the system, and lack of observable states as explained in Section~\ref{sec:moc}. Unlike traditional software development process, where a set of test cases can be executed on a single development computer, functionalities of digital twins cannot be easily validated unless they are tested and compared with the real system, which is often difficult and time consuming. Moreover, it is unrealistic to validate every aspect of physical phenomenon of the twinned system -- should a designer consider all the circuitries, properties of the materials used, geometry, multibody, and etc. of the IMPACT line? How much of the physical system can be abstracted without causing significant discrepancy? For example, after deploying the digital twin for the IMPACT line, discrepancies between the physical plant and the twin might be found based on feedback data collected from the two. In this case, there should be a facility to minimise the error by automatically adjusting the parameters of the twin or even modify the model during runtime. If the error cannot be minimised, the corresponding issue can be reported and possible solutions may be suggested to the user. To realise a faithful digital twin, therefore, the development framework should utilise both modelling (offline) and post-calibration (online) techniques to manage and reduce discrepancy and uncertainty in the digital twin model.

\subsubsection{Interplay with Big Data.}
         
Arguably, big data plays an important role in Industry 4.0. However, there are still many questions need to be answered on how digital twin models interact with big data: what are the gains and losses when employing digital twin based simulations or data-driven approaches for predictive maintenance, decision making, fault detection,  etc. For example, data analytics might detect anomalies in a physical system but may not have a formal model of the plant to draw possible solutions and the root-cause of the problem. On the other hand, a formal model of digital twin itself may not have an ability to deduce useful insights from data generated from the physical plant or high-fidelity simulation. Undoubtedly, there is a potential synergy between the big data and formal modelling of digital twin. Recently, big data topic in digital twin modelling has been recognised in several literature~\cite{qi2018digital,grieves2017digital,zhuang2018digital}. However, a concrete use-case scenarios need to be further investigated.

A concept of Data-driven and Model-based Design (DMD) is discussed in \cite{tripakis2018data} which is an extension of the Model-Based Design (MBD) with elements of data-driven approaches such as machine learning and model learning. The main focus of DMD is to leverage the advances of AI while preserving merits of MBD approaches such as mathematical formalism and analysability. In particular, model learning~\cite{vaandrager2017model} can be useful for building formal models of legacy (black-box) models or for analysing complex systems when there is a lack of tools. Nevertheless, scalability issue when increasing number of states in the system and its applicability for building hybrid system are still in question.

\subsubsection{Integration of Heterogeneous Models.}

There should be a consistent way to combine heterogeneous models to support design of the ``mixed-fidelity'' digital twin. This requires the formal modelling framework that can capture physical behaviour in different levels of abstraction. In case of designing the IMPACT line, it would be easier for a designer to first model a behaviour of the system in coarser level, e.g. flow of workpieces between modules using Petri-Nets or abstracting continuous dynamics using macro states in FSM, and further refine individual modules using high-fidelity models as required. In this case, a consistent method for exchanging information between different models should be developed. For example, when combining two models in Fig.~\ref{fig:rbh-model}, it is crucial that the event $e$ generated from the HA model is always captured in the Petri-Net model. This implies that the models should synchronise whenever the invariant condition $|x_d-x|\geq 0.03$ becomes unsatisfied. Implementation-wise, the FMI standard~\cite{blochwitz2011functional} can be employed for interconnecting the heterogeneous models and implementing a deterministic mechanism for synchronisation and data exchange.





\subsubsection{Time-Critical Systems.}

Some of the cyber-physical production systems are time-critical systems, for example Surface-Mount Technology (SMT) placement equipment. In this case, the worst-case response time for communication between the twin and the physical system must be guaranteed and within a bounded time. Furthermore, the worst-case execution time of the twin and its controller should also be bounded and analysable. Many digital twin architectures found in literature rely on information flow through the cloud and the non time-critical network interfaces, which is not suitable to build time-critical applications. A software architecture that comprises time-critical components are required for certain types of application.

\subsubsection{Security.}

Since digital twins are tightly coupled with physical environment, an attack on a such cyber-physical system may endanger people's safety and result in significant economic loss. Applications interact with digital twins to access sensor data and actuate the physical system when necessary (see Section~\ref{sec:dt-sm}). Since user applications can be a major source of security threat, a secure access control mechanism needs to be implemented for those that access digital twins. In summary, security aspects of digital twin should be studied in the areas of malicious activities detection, cryptography, resiliency to cyber attacks.

\subsection{An Architecture for Digital Twin}\label{sec:arch}

To address the aforementioned issues, the desired capabilities of digital twin need to be clearly defined first which will be the basis for the development of the underlying software/hardware architecture (see Fig.~\ref{fig:cps}). More specifically, we anticipate that through digital twin, users should be able to (1) \textit{control} physical object, (2) \textit{monitor and analyse} data for minimising discrepancy and uncertainties in a model, and (3) \textit{simulate} the physical counterparts by instantiating digital twins in a cyber space. To achieve this, a basic design block will be introduced namely a digital twin component (DTC). It consists of a twin itself, a virtual controller, database for storing cyber and physical data, and a digital twin runtime that manages data collection and synchronisation between the twin and the physical object. Furthermore, DTC should support incremental update of the twin as well as the controller to handle discrepancies and uncertainties as much as possible. Multiple DTCs can be combined to create a bigger system and we plan to adopt Globally Asynchronous Locally Synchronous paradigm for their integration. On the highest level of the architecture, we plan to provide a set of services, for example DTC repository, DTC management (i.e. creation and disposal), analytics, and data access etc., with which more intelligent applications such as prognostic and health management and virtual commissioning can be developed. The digital twin framework therefore will consist of a tool and the architecture that enables the aforementioned capabilities for CPPS.

\section{Conclusions}\label{sec:conclusions}

In this paper, we presented an overview and open research questions in digital twin development for cyber-physical production systems (CPPS). Digital twin is still a relatively new concept that requires more researches in the fields of modelling and integration with other technologies. We propose development of a tool that supports modelling CPPS in various levels of abstraction underpinned by formal mathematical models. The result is a mixed-fidelity digital twin which is a trade-off between accuracy and scalability. In future work, we plan to develop formal semantics for translating different models into an intermediate representation for analysis and code generation and implement a digital twin architecture described in Section~\ref{sec:arch} that enables control, monitoring, and simulation of physical plant in the cyber space.

\bibliographystyle{splncs04}
\bibliography{bibs}

\begin{thebibliography}{10}
\providecommand{\url}[1]{\texttt{#1}}
\providecommand{\urlprefix}{URL }
\providecommand{\doi}[1]{https://doi.org/#1}

\bibitem{alam2017c2ps}
Alam, K.M., El~Saddik, A.: {C2PS: A Digital Twin Architecture Reference Model
  for the Cloud-based Cyber-Physical Systems}. IEEE Access  \textbf{5},
  2050--2062 (2017)

\bibitem{alur1991theory}
Alur, R., Dill, D.: {A Theory of Timed Automata}. In: Workshop/School/Symposium
  of the REX Project (Research and Education in Concurrent Systems). pp.
  45--73. Springer (1991)

\bibitem{berry1992esterel}
Berry, G., Gonthier, G.: {The ESTEREL Synchronous Programming Language: Design,
  Semantics, Implementation}. Science of computer programming  \textbf{19}(2),
  87--152 (1992)

\bibitem{bliudze2017rigorous}
Bliudze, S., Furic, S., Sifakis, J., Viel, A.: {Rigorous Design of
  Cyber-Physical Systems}. Software \& Systems Modeling pp. 1--24 (2017)

\bibitem{blochwitz2011functional}
Blochwitz, T., Otter, M., Arnold, M., Bausch, C., Elmqvist, H., Junghanns, A.,
  Mau{\ss}, J., Monteiro, M., Neidhold, T., Neumerkel, D., et~al.: {The
  Functional Mockup Interface for Tool Independent Exchange of Simulation
  Models}. In: Proceedings of the 8th International Modelica Conference; March
  20th-22nd; Technical Univeristy; Dresden; Germany. pp. 105--114. No.~063,
  Link{\"o}ping University Electronic Press (2011)

\bibitem{brooks2008heterogeneous}
Brooks, C., Lee, E.A., Liu, X., Neuendorffer, S., Zhao, Y., Zheng, H.,
  Bhattacharyya, S.S., Cheong, E., Davis, I., Goel, M., et~al.: {Heterogeneous
  Concurrent Modeling and Design in Java (Volume 1: Introduction to Ptolemy
  II)}. Tech. rep., California Univ. Berkeley Dept. Of Electrical Engineering
  and Computer Science (2008)

\bibitem{fehling1991concept}
Fehling, R.: {A Concept of Hierarchical Petri Nets with Building Blocks}. In:
  International Conference on Application and Theory of Petri Nets. pp.
  148--168. Springer (1991)

\bibitem{fritzson2010principles}
Fritzson, P.: {Principles of Object-Oriented Modeling and Simulation with
  Modelica 2.1}. John Wiley \& Sons (2010)

\bibitem{ge-wind-farm}
{Generic Electric}: {Digital Wind Farm},
  \url{https://www.ge.com/renewableenergy/wind-energy/technology/digital-wind-farm}

\bibitem{glaessgen2012digital}
Glaessgen, E., Stargel, D.: {The Digital Twin Paradigm for Future NASA and US
  Air Force Vehicles}. In: 53rd AIAA/ASME/ASCE/AHS/ASC Structures, Structural
  Dynamics and Materials Conference 20th AIAA/ASME/AHS Adaptive Structures
  Conference 14th AIAA. p.~1818 (2012)

\bibitem{grieves2017digital}
Grieves, M., Vickers, J.: {Digital Twin: Mitigating Unpredictable, Undesirable
  Emergent Behavior in Complex Systems}. In: Transdisciplinary Perspectives on
  Complex Systems, pp. 85--113. Springer (2017)

\bibitem{henzinger2000theory}
Henzinger, T.A.: {The Theory of Hybrid Automata}. In: Verification of Digital
  and Hybrid Systems, pp. 265--292. Springer (2000)

\bibitem{hoare1998unifying}
Hoare, C.A.R., Jifeng, H.: {Unifying Theories of Programming}, vol.~14.
  Prentice Hall Englewood Cliffs (1998)

\bibitem{jensen2015colored}
Jensen, K., Kristensen, L.M.: {Colored Petri Nets: A Graphical Language for
  Formal Modeling and Validation of Concurrent Systems}. Communications of the
  ACM  \textbf{58}(6),  61--70 (2015)

\bibitem{kagermann2013recommendations}
Kagermann, H., Helbig, J., Hellinger, A., Wahlster, W.: {Recommendations for
  Implementing the Strategic Initiative INDUSTRIE 4.0: Securing the Future of
  German Manufacturing Industry; Final Report of the Industrie 4.0 Working
  Group}. Forschungsunion (2013)

\bibitem{larsen2016integrated}
Larsen, P.G., Fitzgerald, J., Woodcock, J., Fritzson, P., Brauer, J., Kleijn,
  C., Lecomte, T., Pfeil, M., Green, O., Basagiannis, S., et~al.: {Integrated
  Tool Chain for Model-based Design of Cyber-Physical Systems: The INTO-CPS
  Project}. In: Modelling, Analysis, and Control of Complex CPS (CPS Data),
  2016 2nd International Workshop on. pp.~1--6. IEEE (2016)

\bibitem{lee2008cyber}
Lee, E.A.: {Cyber Physical Systems: Design Challenges}. In: Object Oriented
  Real-Time Distributed Computing (ISORC), 2008 11th IEEE international
  symposium on. pp. 363--369. IEEE (2008)

\bibitem{li2017dynamic}
Li, C., Mahadevan, S., Ling, Y., Wang, L., Choze, S.: {A Dynamic Bayesian
  Network Approach for Digital Twin}. In: 19th AIAA Non-Deterministic
  Approaches Conference. p.~1566 (2017)

\bibitem{lopez2017software}
Lopez, F., Shao, Y., Mao, Z.M., Moyne, J., Barton, K., Tilbury, D.: {A
  Software-Defined Framework for the Integrated Management of Smart
  Manufacturing Systems}. Manufacturing Letters  (2017)

\bibitem{ansys-2017-dt-pump}
MacDonald, C., Dion, B., Davoudabadi, M.: {Creating a Digital Twin for a Pump}.
  ANSYS Advantage Issue 1 p.~8 (2017)

\bibitem{magargle2017simulation}
Magargle, R., Johnson, L., Mandloi, P., Davoudabadi, P., Kesarkar, O.,
  Krishnaswamy, S., Batteh, J., Pitchaikani, A.: {A Simulation-Based Digital
  Twin for Model-Driven Health Monitoring and Predictive Maintenance of an
  Automotive Braking System}. In: Proceedings of the 12th International
  Modelica Conference, Prague, Czech Republic, May 15-17, 2017. pp. 35--46.
  No.~132, Link{\"o}ping University Electronic Press (2017)

\bibitem{malik2010systemj}
Malik, A., Salcic, Z., Roop, P.S., Girault, A.: {SystemJ: A GALS Language for
  System Level Design}. Computer Languages, Systems \& Structures
  \textbf{36}(4),  317--344 (2010)

\bibitem{website:mam-dt}
MarketsandMarkets: {Digital Twin Market by End User (Aerospace \& Defense,
  Automotive \& Transportation, Home \& Commercial, Electronics \&
  Electricals/Machine Manufacturing, Energy \& Utilities, Healthcare, Retail \&
  Consumer Goods), and Geography} (August 2017),
  \url{http://www.reportsnreports.com/reports/1175159-digital-twin-market-by-end-user-aerospace-defense-automotive-transportation-home-commercial-electronics-electricals-machine-manufacturing-energy-utilities-healthcare-retail-consumer-goods-and-ge-st-to-2023.html}

\bibitem{moreno2017virtualisation}
Moreno, A., Velez, G., Ardanza, A., Barandiaran, I., de~Infante, {\'A}.R.,
  Chopitea, R.: {Virtualisation Process of a Sheet Metal Punching Machine
  within the Industry 4.0 Vision}. International Journal on Interactive Design
  and Manufacturing (IJIDeM)  \textbf{11}(2),  365--373 (2017)

\bibitem{negri2017review}
Negri, E., Fumagalli, L., Macchi, M.: {A Review of the Roles of Digital Twin in
  CPS-based Production Systems}. Procedia Manufacturing  \textbf{11},  939--948
  (2017)

\bibitem{peterson1981petri}
Peterson, J.L.: Petri Net Theory and the Modeling of Systems. Prentice Hall
  PTR, Upper Saddle River, NJ, USA (1981)

\bibitem{potok2018sdcworks}
Potok, M., Chen, C.Y., Mitra, S., Mohan, S.: {SDCWorks: A Formal Framework for
  Software Defined Control of Smart Manufacturing Systems}. In: Proceedings of
  the 9th ACM/IEEE International Conference on Cyber-Physical Systems. pp.
  88--97. IEEE Press (2018)

\bibitem{qi2018digital}
Qi, Q., Tao, F.: {Digital Twin and Big Data Towards Smart Manufacturing and
  Industry 4.0: 360 Degree Comparison}. IEEE Access  \textbf{6},  3585--3593
  (2018)

\bibitem{ramchandani1974analysis}
Ramchandani, C.: {Analysis of Asynchronous Concurrent Systems by Petri Nets}.
  Tech. rep., MASSACHUSETTS INST OF TECH CAMBRIDGE PROJECT MAC (1974)

\bibitem{scaglioni2018towards}
Scaglioni, B., Ferretti, G.: {Towards Digital Twins Through Object-Oriented
  Modelling: A Machine Tool Case Study}. IFAC-PAPERSONLINE pp. 613--618 (2018)

\bibitem{schluse2018experimentable}
Schluse, M., Priggemeyer, M., Atorf, L., Rossmann, J.: {Experimentable Digital
  Twins—Streamlining Simulation-Based Systems Engineering for Industry 4.0}.
  IEEE Transactions on Industrial Informatics  \textbf{14}(4),  1722--1731
  (2018)

\bibitem{siemens-bausch}
Siemens: {Digital twins bring real-life success},
  \url{https://www.siemens.com/global/en/home/markets/machinebuilding/references/bausch-stroebel.html}

\bibitem{tao2018digital}
Tao, F., Cheng, J., Qi, Q., Zhang, M., Zhang, H., Sui, F.: {Digital Twin-Driven
  Product Design, Manufacturing and Service with Big Data}. The International
  Journal of Advanced Manufacturing Technology  \textbf{94}(9-12),  3563--3576
  (2018)

\bibitem{tao2017digital}
Tao, F., Zhang, M.: {Digital Twin Shop-Floor: A New Shop-Floor Paradigm Towards
  Smart Manufacturing}. IEEE Access  \textbf{5},  20418--20427 (2017)

\bibitem{tripakis2018data}
Tripakis, S.: {Data-Driven and Model-based Design}. In: 2018 IEEE Industrial
  Cyber-Physical Systems (ICPS). pp. 103--108. IEEE (2018)

\bibitem{vaandrager2017model}
Vaandrager, F.: {Model Learning}. Communications of the ACM  \textbf{60}(2),
  86--95 (2017)

\bibitem{wang2017human}
Wang, X.V., Kem{\'e}ny, Z., V{\'a}ncza, J., Wang, L.: {Human--Robot
  Collaborative Assembly in Cyber-Physical Production: Classification Framework
  and Implementation}. CIRP annals  \textbf{66}(1), ~5--8 (2017)

\bibitem{zhuang2018digital}
Zhuang, C., Liu, J., Xiong, H.: {Digital Twin-based Smart Production Management
  and Control Framework for the Complex Product Assembly Shop-Floor}. The
  International Journal of Advanced Manufacturing Technology  \textbf{96}(1-4),
   1149--1163 (2018)

\end{thebibliography}
\end{document}